\documentclass[journal,10pt,twocolumn,letter]{IEEEtran}

\usepackage{cite}
\usepackage{graphicx}
\usepackage{lipsum}
\usepackage[intlimits]{amsmath}
\usepackage{flushend}
\usepackage{color}
\usepackage{amsmath}
\usepackage{amssymb}
\usepackage{citesort}

\begin{document}

\title{Analysis of Network Coding Schemes for Differential Chaos Shift Keying Communication System}

\author{Georges Kaddoum$^{*}$,~\IEEEmembership{Member,~IEEE,} and Mohammed El-Hajjar,~\IEEEmembership{Member, IEEE}   

\thanks{G. Kaddoum is with University of Qu\'{e}bec, ETS, LaCIME Laboratory, 1100 Notre-Dame west, H3C 1K3, Montreal, Canada (e-mail: georges.kaddoum@etsmtl.ca). 
M. El-Hajjar is with the ECS, Faculty of Physical Sciences and Engineering, University of Southampton, Southampton SO17 1BJ, United Kingdom (e-mail: meh@ecs.soton.ac.uk).

* This work has been supported by the NSERC discovery grant $435243-2013$.}
}


\maketitle

\begin{abstract}
In this paper we design network coding schemes for Differential Chaos Shift Keying (DCSK) modulation. In this work, non-coherent chaos-based communication system is used due to its simplicity and robustness to multipath propagation effects, while dispensing with any channel state information knowledge at the receiver. We propose a relay network using network coding and DCSK, where we first present a Physical layer Network Coding (PNC) scheme with two users, $\mathcal{A}$ and $\mathcal{B}$, sharing the same spreading code and bandwidth, while synchronously transmitting their signals to the relay node $\mathcal{R}$. We show that the main drawback of this design in multipath channels is the high level of interference in the resultant signal, which severely degrades the system performance.  Hence, in order to address this problem, we propose two coding schemes, which separate the users' signals in the frequency or the time domains. We show also in this paper that the performance of the Analog Network Coding (ANC) with DCSK modulation suffers from the same interference problem as the PNC scheme.
We present the analytical bit error rate performance for multipath Rayleigh fading channel for the different scenarios and we analyse these schemes in terms of complexity, throughput and link spectral efficiency.
\end{abstract}



\section{Introduction}
\IEEEPARstart{H}{igh} throughput and low interference are important design criteria for any wireless network. Recently, there has been a major benefit of employing network coding in cooperative communications \cite{Haj1,Haj2,5896023}, such as Straightforward Network Coding (SNC), Analog Network Coding (ANC) and Physical layer Network Coding (PNC), where it has been shown that higher throughput and lower interference for simultaneously arriving signals coming from multiple nodes can be achieved~\cite{soun11,Katti07,Zhan06,Fragouli_2006,5699899,5288500}.
Typically, a PNC, SNC or ANC communication scheme consists of a pair of users nodes $\mathcal{A}$ and $\mathcal{B}$, that communicate with each other via a relay node $\mathcal{R}$. 
In a PNC scheme, the relay received the signals from the two users $\mathcal{A}$ and $\mathcal{B}$ and converts it into a network-coding message, such as bit-wise XORing the two users' messages before broadcasting the network-coded message. It is essential that the network coding operation carried out by the relay is reversible and is also known by the users. Then, each user can remove its own data message in order to detect the message sent by the other node. This method for the PNC scheme requires only two time slots to exchange one message from one node to another. It should be mentioned that in network coding schemes, a time slot for a given source or relay is defined as the required time to transmit its data packet \cite{5699899}. 

Furthermore, in the ANC schemes, the relay simply transmits the received signal superimposed to the pair of user nodes~\cite{Katti07}. The linear mapping is naturally done in the physical channel and the signals are decoded by the user nodes. In this scheme, due to the lack of noise cancellation at the relay side, the noise is forwarded with the superimposed signals, which leads to the degradation of the system performance. 

Contrary to the ANC and PNC schemes, SNC technique does not need any precise user synchronization and just a network synchronization is required. The SNC is a non-physical-layer network coding and requires three time slots to exchange a data message between the two user nodes~\cite{Fragouli_2006}. The first and the second time slots are used by user $\mathcal{A}$ and user $\mathcal{B}$ respectively in order to transmit their signals to the relay. The signals sent by the two users are separately decoded and mapped by the relay before forwarding the combined signal to the two users in the third time slot. 

Recently, there has been research focus on chaotic signals due to their advantageous characteristics that can be used in the field of wireless communications~\cite{Vitali2006,Kur05,Val12,Fan13,Fan13i,Wei11,chi08,kad13arx,6486530,4631471}. This is principally associated with the inherent wideband characteristics of these signals, which make them well suited for spread-spectrum modulation systems~\cite{Lau03,Kur05,Manoharan2014}. Chaotic modulations have similar advantages to all other spread-spectrum modulations, including the mitigation of fading channels and jamming resistance \textit{exempli gratia}. Furthermore, the low probability of interception (LPI) in chaotic signals~\cite{Yu2005,6179320} and excellent correlation properties \cite{Manoharan2014} allow them to be one of the natural candidates for military communication scenarios. 

Various digital chaos-based communication schemes have been evaluated and analysed including coherent chaos-shift-keying (CSK)~\cite{Lau03,Kad09ieee,Kad08cs}, chaos-based DS-CDMA~\cite{Vitali2006,Kur05,Val12} and non-coherent Differential Chaos Shift Keying (DCSK)~\cite{Fan13,Fan13i,Wei11,Jing10,kadiet14}. In CSK and chaos-based DS-CDMA, chaotic sequences are used to spread data signals instead of conventional spreading codes used in DS-CDMA. One of the main drawbacks of coherent chaos-based communication systems, such as the CSK, is that they require chaotic synchronization which is non-trivial. For instance, the chaotic synchronization proposed by Pecora and Carroll in~\cite{Pec97} is still practically impossible to achieve in a noisy environment and hence the coherent system can not be used in realistic applications.

The research presented in this paper is based on one of the most commonly used non-coherent chaotic modulation techniques known as differential chaos shift keying (DCSK). In this system, each bit duration is split into two equal slots, where the first slot is allocated to the reference chaotic signal, while the second slot is used to transmit either the reference signal or its inverted version, depending on the bit being sent. The analytical performance derivation of DCSK communication systems has been presented in~\cite{kad12wpc} for fading channels and in~\cite{Fan13,Fan13i,Wei11,Jing10,kadiet14} for cooperative and Multiple-Input Multiple-Output (MIMO) schemes. Moreover, DCSK has been used in several systems in the literature, such as the design of new ultra-wideband systems based on DCSK, Multi Carrier DCSK (MC-DCSK) and Frequency Modulated DCSK (FM-DCSK) schemes \cite{Cima07,chi08,Min10,6560492}.

\subsubsection*{\textbf{Motivation}}

The DCSK modulation is chosen as candidate for network coding schemes due to its various advantageous. In fact, the chaotic signal generation and synchronization are not required at the receiver side of DCSK demodulator ~\cite{Fan13,Fan13i,Wei11,Jing10,kadiet14} which makes this system easy to implement~\cite{Kadgnu}. On the other hand, the common points between DCSK and differential phase shift keying (DPSK) modulations is that both are non-coherent schemes and do not require channel state information at the receiver to recover the transmitted data~\cite{Rap96,Xia04,5648457}. However, DCSK system is more robust to multipath fading environment than the DPSK scheme~\cite{Xia04} and is suitable for Ultra-Wide band (UWB) applications~\cite{kad13arx,chi08,Chen13,Xia04,5648457}.

 \subsubsection*{\textbf{Related work}}
In~\cite{Kulhandjian12,Shu12,5200989,5198836,6952033}, ANC and PNC schemes have been proposed for spread spectrum systems and their performance was compared in~\cite{5198836} for different scenarios and wireless channels.
However, it should be highlighted that the proposed schemes with coherent detection assume perfect knowledge of the channel state information (CSI), where the assumption of having perfect CSI at the relay or user ends is impractical~\cite{eps342888}. Therefore, a DCSK modulation is preferred to avoid the strong unrealistic assumption of the perfect channel knowledge. In~\cite{Kulhandjian12,Shu12}, each user in the system is assigned a specific spreading sequence, where the destination node applies the spreading sequence of the desired source node in order to decode its message. In this case, the decoder requires a prior knowledge of the list of all users with their specific spreading sequences in order to decode the transmitted data without facing any multi-user interference.
Recently, an analog network scheme for Multi-User Multi-Carrier Differential Chaos Shift Keying Communication System was proposed in \cite{Kad_wir}. This system support the multi-user transmission but is designed for ultra wide band communications. The required band for this MC-DCSK system is $M$ times the band of the conventional DCSK one where $M$ is the number of transmitted bits.Therefore, for limited available bandwidth, a DCSK system can be preferred for such application.  

\subsubsection*{\textbf{Contributions}}
This paper proposes novel network coding schemes for non-coherent chaos-based spread spectrum systems. The structure of this network is considered as a full duplex wireless system with one single relay, where each user node uses DCSK modulation. The motivation to use DCSK is associated with its offered advantages described above. 

The novel contribution of this paper are summarized as follows:
\begin{enumerate}
\item The first proposed scheme, which is denoted as scheme $1$ in this work, is physical layer network coding DCSK (PNC-DCSK). In this scheme, users $\mathcal{A}$ and $\mathcal{B}$ share the same spreading code and bandwidth and transmit their signals to the relay synchronously. The relay then decodes and maps the combined received signal and forwards the resultant signal to the users. We present the performance analysis of scheme $1$, where we show that it suffers from a strong interference originating from the cross product of the user's signals at the relay's correlator.
\item Hence, in order to address the interference problem of scheme $1$, two other coding schemes, denoted as scheme $2$ and scheme $3$, are proposed in section IV. In these schemes, the user signals are separated in the first phase of cooperation in time domain for scheme $2$ and in frequency domain for scheme $3$. This paper shows that both schemes reduce the interference as well as enhance the bit error rate performance of the system. Scheme $2$ takes advantage of multiplexing in time domain and is equivalent to SNC scheme, whereas scheme $3$ is equivalent to PNC-DCSK, requiring twice the conventional PNC scheme's bandwidth due to multiplexing in frequency domain.
\item We analyse the corresponding interference levels and derive the analytical end-to-end bit error ratio expressions for scheme $2$ and scheme $3$ over multipath fading channel for different scenarios. Finally, we analyse the different BER performances of the PNC, ANC, SNC schemes, the throughput and the link spectral efficiency of the two systems. To the best of author’s knowledge, there is no previous publications investigating on the problem of DCSK modulation with network coding schemes.

\end{enumerate}
%
 

 The remainder of this paper is organized as follows. In Section II, the DCSK spread-spectrum communication system is briefly presented. Section III is dedicated to the design and the performance analysis of the proposed PNC-DCSK scheme (i.e scheme $1$). Section IV presents the design of the scheme $2$ and $3$. Additionally in this section, the analytical bit error rate (BER), the throughput and the link spectral efficiency expressions of the two schemes are derived under multipath Rayleigh fading channels. Finally in Section V, the simulation results are presented and the obtained analytical expressions of this work are evaluated with some concluding remarks. 

\section{DCSK Communication System} 
\label{dcsk_demod}
The broadband nature of chaotic signals and their good correlation properties as well as the ease with which they can be generated have made them special type of signals, which can be advantageously used to design and implement spread spectrum communication systems. 

\begin{figure}[htbp]
\centering 
\includegraphics[width=1.0\linewidth]{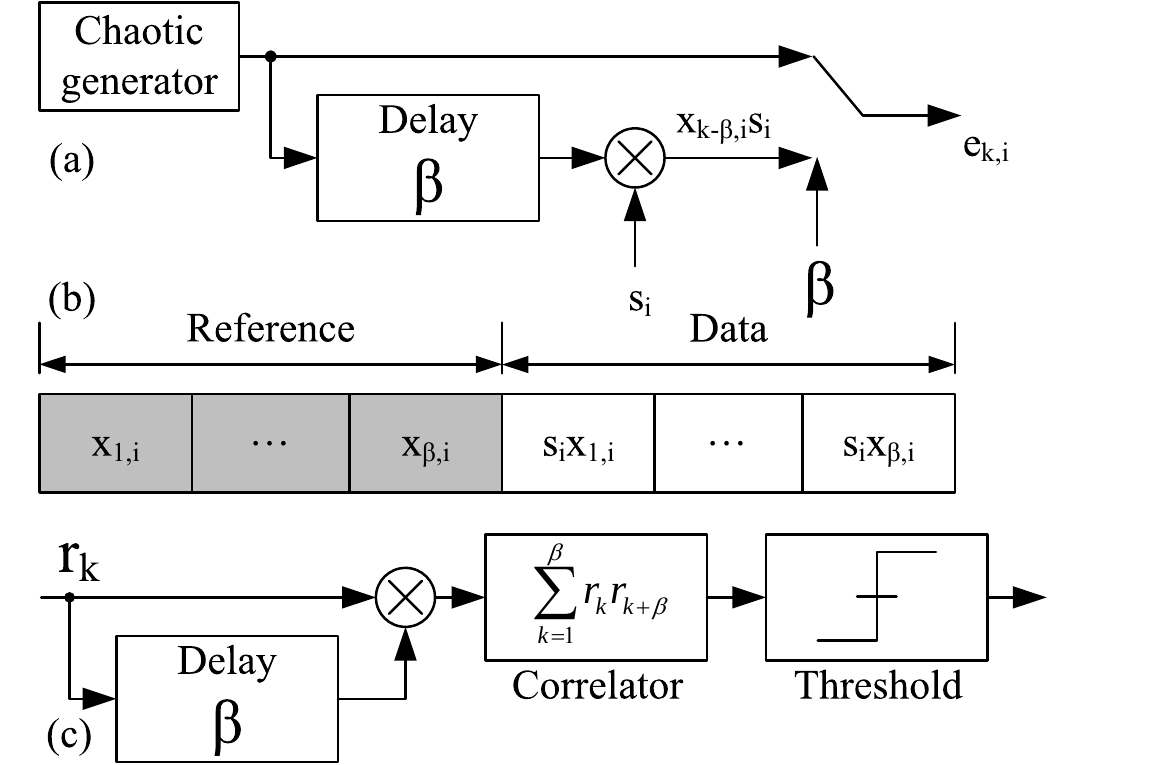}
\caption {Block diagram of the general structure of the DCSK communication system; (a) The DCSK transmitter, (b) The DCSK frame, and (c) The DCSK receiver. 
\label{DCSKSYS}} 
\end{figure} 

Fig.~\ref{DCSKSYS} summarizes the structure of a single user DCSK communication system. At the transmitter side, each bit $s_{i}=\{-1, +1\}$ is represented by two consecutive sets of chaotic signals: the reference signal followed by the data sequence. Depending on whether the sent bit is $+1$ or $-1$, the reference sequence or its inverted version is used as the data bearing sequence. During the $i^{th}$ bit duration, the output of the transmitter $e_{i,k}$ shown in Fig.~\ref{DCSKSYS} can be given by
\begin{equation}
\label{DCSKTransmitterOutput}
e_{i,k} =  
  \begin{cases}
    x_{i,k}            & \quad \textrm{for}\,\,  1< k \leq \beta \\
    s_{i}x_{i,k-\beta} & \quad \textrm{for}\,\,  \beta < k \leq 2\beta
  \end{cases}\,\,,
\end{equation}
where $x_{k}$ and $x_{k-\beta}$ denote the reference sequence and its delayed version, respectively. Here, $\beta$ is an integer and 2$\beta$ denotes the spreading factor which is determined by the number of chaotic samples sent for each bit as shown in Fig.~\ref{DCSKSYS}(b). 

In order to demodulate the transmitted bits at the receiver side, the received signal $r_{k}$ is correlated with its delayed version $r_{k + \beta}$ and summed over a half bit duration $T_{b}/2$, where $T_{b}= 2\beta T_{c}$ and $T_{c}$ denotes the chip time. This demodulation process can be performed without any need for channel state information at the receiver side, which is a benefit of the insertion of the reference signal after each symbol. Finally, the sign of the correlator output is then computed to estimate the received bits, as shown in Fig.~\ref{DCSKSYS}(c).

In this work, the second-order Chebyshev polynomial function (CPF) is employed to generate chaotic sequences due to its simplicity and good performance~\cite{769472}. For simplicity, the chip time is set to one, i.e. $T_c=1$, hence the sequence $x$ can be obtained as
\begin{equation}\label{cpf}
 x_{k+1}=1-2x_{k}^{2} \,\cdot
\end{equation}

The variance of the normalized chaotic map with zero is equal to one, i.e. $\textrm{Var}(x)= \textrm{E}[x^2]=1$, where $\textrm{E}[\: .\: ]$ denotes the expected value operation.

\section{pnc-dcsk communication scheme} 
\label{sect:scheme}
%
\begin{table*}[htb!]
\small 
\caption{First PNC-DCSK mapping scheme} 
\centering 
\begin{tabular}{|r| r| r| c| c| r| c| c| c |} 
\hline\hline 
 $e_{\mathcal{A}}\,\,\,\,\,\,\,$    & $e_{\mathcal{B}}\,\,\,\,\,\,\,\,$  &  $r_{\mathcal{R}}\,\,\,\,\,\,\,\,$ &  $s_{\mathcal{R},D}$ &$s_{\mathcal{R},M}$ & $e_\mathcal{R}\,\,\,\,\,\,\,\,$ &  $s_{\mathcal{(A,B)},D}$  &   $\begin{array}{l} s_{\mathcal{(A,B)},M} =\\  \mid s_{\mathcal{(A,B)},D} +s_{B} \mid \end{array}$&   $\begin{array}{l} \hat s_{\mathcal{A}}=\\s_{\mathcal{(A,B)},M}- 1\end{array}$\\[0.05ex] \hline\hline
    $[x,1x]$    &   $[x,1x]$   &  $[2x,2x]$&  2  & $\,\,1$ & $[x,1x]$ & $\,\,1$ &  2 & $1$ \\ \hline
    $[x,1x]$   &  $[x,-1x]$   & $[2x,0x]$ &  0 & -1 & $[x,-1x]$ & -1 &   2 &  $1$ \\ \hline
    $[x,-1x]$ &  $[x,1x]$    & $[2x, 0x]$ &  0 & -1 & $[x,-1x]$ & -1 &  0 & $-1\,\,\,\,\,$ \\ \hline
    $[x,-1x]$ &  $[x,-1x]$   & $[2x,-2x]$&  -2  & $\,\,1$ & $[x,1x]$ & $\,\,1$ & $ \,\,\,\, 0\,\,\,\,\,$ &$-1\,\,\,\,\,$  \\ \hline
  \end{tabular}\label{map-PNC_DCSK}
\end{table*}

\begin{figure}[htbp]
\centering 
\includegraphics[width=0.8\linewidth]{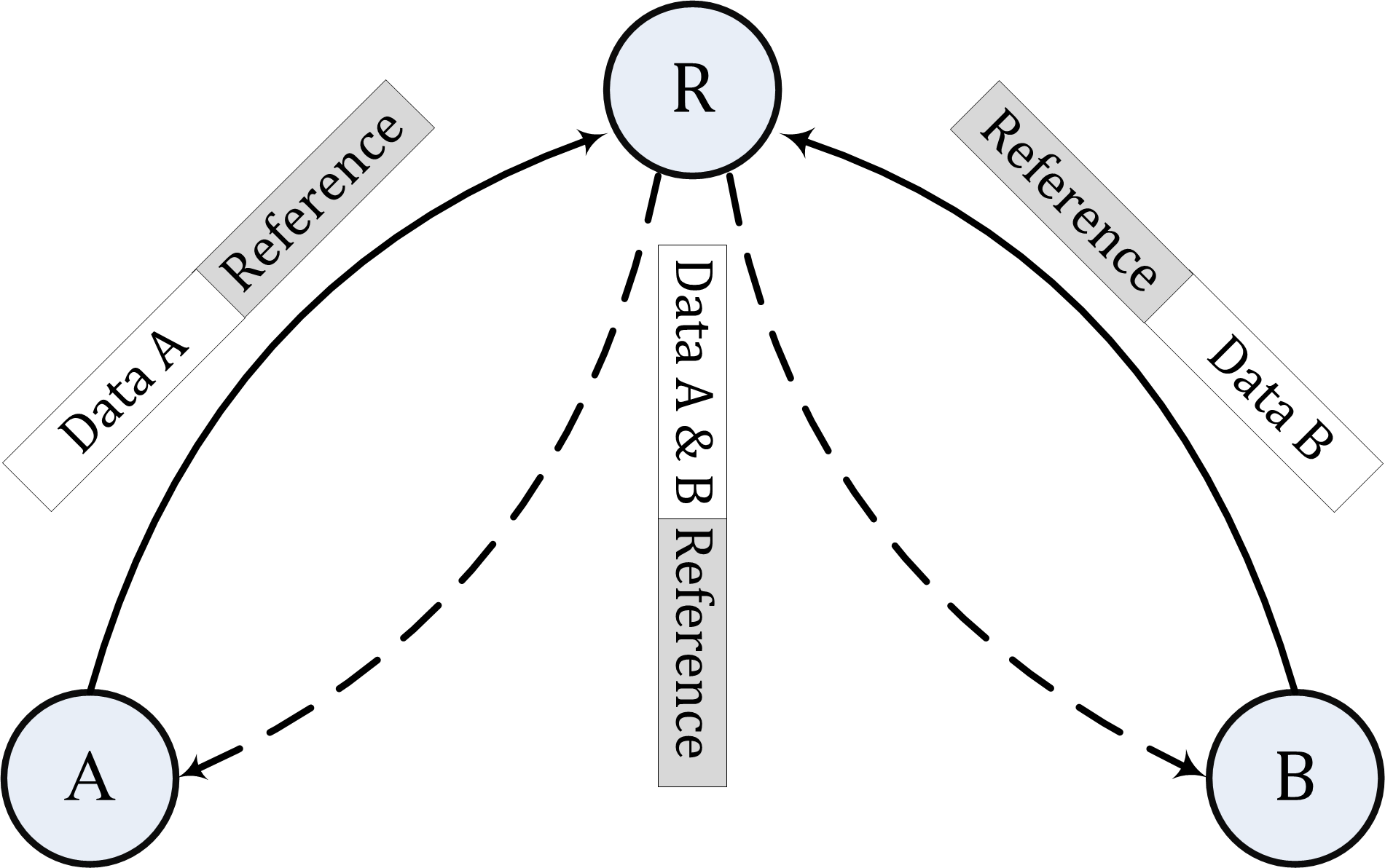}
\caption {Block diagram of the general structure of network coding DCSK communication system.  
\label{PNC_DCKS_Topology}} 
\end{figure}

Fig.~\ref{PNC_DCKS_Topology} illustrates the topology of the proposed PNC-DCSK system, which we refer to as scheme $1$. The users synchronization process is carried out by the relay sending a \emph{clear-to-send} (CTS) message in response to the \emph{Request to send} (RTS) message sent by the user nodes. For example, in the system shown in Fig.~\ref{PNC_DCKS_Topology}, we consider two nodes $\mathcal{A}$ and $\mathcal{B}$, who want to communicate with each other and they want to transmit their message frame to the relay simultaneously. 

In this scheme, the user nodes utilise the same spreading sequence, i.e. chaotic signal, in their data frame, which is sent to the relay node. As a consequence of using the same spreading sequences, the number of the chaotic codes present in the channel is reduced to one and thus lower multiple access interference is experienced by each user and also the power of the useful resultant signal is boosted. The users' transmitted data frames are then superimposed in the wireless channel, when they are received by the relay node. The relay then despreads and decodes the received combined signal. The decoded symbols are then mapped by the relay before modulating and broadcasting them to the user nodes. It should be noted that the relay uses of the same modulation scheme as the nodes, i.e. DCSK modulation.

The PNC-DCSK communication protocol is summarized in table~\ref{map-PNC_DCSK}, where $s_{A}$ and $s_{B}$ represent the data messages of user nodes $\mathcal{A}$ and $\mathcal{B}$, respectively. Furthermore, $e_{A}$, $e_{B}$ and $e_{R} $ refer to the transmitted signals from the user pair to the relay node $\mathcal{R}$ and from the relay node to the user nodes, respectively. Additionally,, the transmitted signal represented by $[x,sx]$ expresses the DCSK frame, where the first term represents the reference sequence $x$ and the second term denotes the symbol $s_{i}=\{-1, +1\}$ multiplied by the reference sequence $x$. Additionally, $s_{\mathcal{R},D}$, $s_{\mathcal{R},M}$ are the decoded and mapped symbols of the received superimposed signal at the relay side. 

As can be seen from table~\ref{map-PNC_DCSK}, the mapping function at the relay converts the decoded symbols of the received signal $r_\mathcal{R}$ by mapping $2$'s to $1$ and $0$'s to $-1$. Using the DCSK modulator at the relay end, the mapped bits $s_{\mathcal{R},M}$ are modulated into a DCSK frame $e_{R}$. Once the relay broadcasts its final DCSK data frame to the two nodes, the received information frame is decoded and demapped by the two users. For instance, from the user $\mathcal{B}$ point of view, the received signal is first despread and decoded. The decoded symbols $s_{\mathcal{(A,B)},D}$ are then demapped into $s_{\mathcal{(A,B)},M}$ by computing the absolute value of $\vert s_{\mathcal{(A,B)},D} +  s_{B} \vert $. Finally, using the same mapping function carried out by the relay, the data transmitted by user $\mathcal{A}$ can be recovered by user $\mathcal{B}$ which is given by $ \hat s_{A}$ in the table. 

\subsection{\textbf{Channel model}}
\begin{figure}[htb!]
\centering
\includegraphics[width=\linewidth]{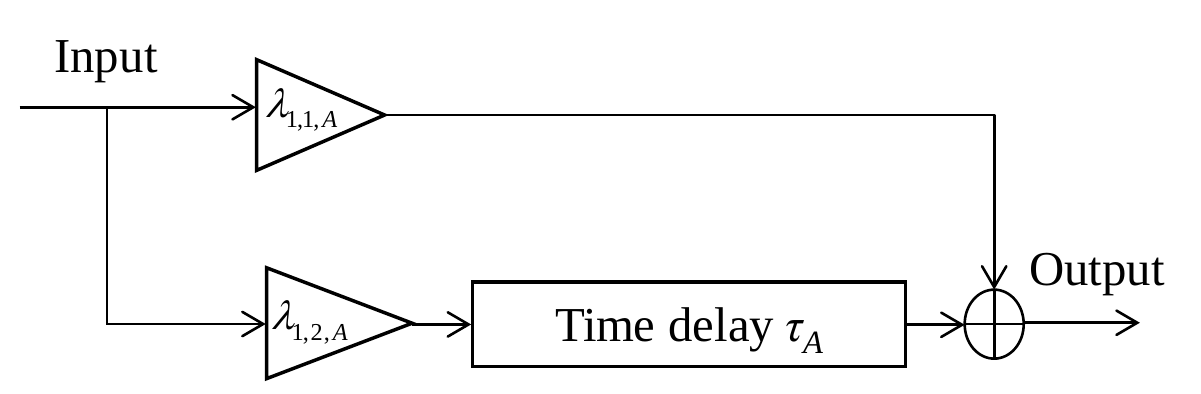}
\caption{Multipath Rayleigh fading model of user $\mathcal{A}$ of the first phase of relaying. \label{chan_mod}}
\end{figure}

In this paper, we use a commonly used channel called two-ray Rayleigh fading channel~\cite{Xia04,Rap96,Chen13}. As shown in Fig.~\ref{chan_mod}, the multipath Rayleigh fading channel with two independent paths is considered for each user. The channel coefficient and time delay between the two rays in the first phase of transmission for user $\mathcal{A}^{th}$ are denoted by $\lambda_{1,2,A}$ and $\tau_{A}$, respectively. It should be noted that the channel coefficients follow the Rayleigh distribution and in this work, they are considered to be constant, i.e., static, over the DCSK frame period $T=2\beta T_c$, while they vary from one frame to another. 

Therefore, the probability density function of the channel coefficient $\lambda$ in this case can be given as the following
\begin{equation}
f(\lambda \vert \, \sigma ) = \frac{_\lambda}{{\sigma ^2 }}e^{ - \frac{{_\lambda^2 }}{{2\sigma ^2 }}}  \:, \:\: \:\: \:\: \: \lambda \ge 0,
\end{equation}
where $\sigma >0$ is the scaling factor of the distribution representing the root mean square value of the received voltage signal before envelope detection.

In this paper, the largest multipath time delay is assumed to be shorter than the bit duration, i.e. $0 <\tau << 2\beta T_c$. Hence, the different time delay values are considered constant during each phase as they do not cause any significant difference, in other words the intersymbol interference (ISI) can be neglected. Therefore, as a consequence of using DCSK modulator in this network, this scheme benefits from high resistance to multipath interference. Moreover, this system does not require any CSI knowledge at the receiver for decoding the data, since the reference plays the role of pilot signal~\cite{Xia04}. 

\subsection{\textbf{Performance analysis of PNC-DCSK system}}
In this section, we analyse the performance of the PNC-DCSK system, where we show that the current form of this scheme, which we refer to as scheme $1$, has high levels of interference, which degrades the system performance dramatically.

As mentioned in Section~\ref{sect:scheme}, in the first phase of relaying, users $\mathcal{A}$ and $\mathcal{B}$ transmit their signals to the relay $\mathcal{R}$ by using the same reference, i.e. spreading sequence. Then, The relay decodes the resultant received signal followed by modulating the decoded bits into a DCSK frame signal and forwards this to the users in the second phase of relaying. 

After receiving the RTS message from the relay, the two users transmit their signals to the relay simultaneously. The received signal at the relay $r_{R}(t)$ can be formulated as follow
\begin{equation}
\begin{array}{l}
 r_R (t) = \lambda _{1,1,A} e_A (t ) + \lambda _{1,2,A}e_A (t - \tau _{A} )  \\ 
\quad  \quad \quad\,\, +  \lambda _{1,1,B} e_B (t ) + \lambda _{1,2,B}e_B (t - \tau _{B} )+ n(t) , 
 \end{array}
\end{equation}
where $\tau$ represents the path delay and  $\lambda_{1,l,A}$ denotes the channel coefficient in the first phase of relaying corresponding to the $l^{th}$ path of user $\mathcal{A}$. Additionally, $e(t)$ is the emitted signal and $n(t)$ represents additive white Gaussian noise (AWGN) with zero mean and variance of $N_0/2$.
Then, to demodulate the transmitted signals by the users, the received signal $r_{R,k}$ at the relay is correlated by its delayed version $r_{R,k + \beta}$ and summed over a half bit duration $T_{b}$ before being compared with zero.

The decision variable for a given $i^{th}$ bit transmitted from $\mathcal{A}$ and $\mathcal{B}$ can be formulated as:
 \begin{equation} \label{dec_relay0}
\begin{array}{l}
 D_{R,i} = \left( {\lambda _{1,1,A}^{} s_{i,A} x_k^{}  + \lambda _{1,1,B}^{} s_{i,B} x_k^{}   } \right. \\ 
 \quad\quad\quad \left. {+ \lambda _{1,2,A}^{} s_{i,A} x_{k - \tau _A }^{} + \lambda _{1,2,B}^{} s_{i,B} x_{k - \tau _B }^{}  + n_k } \right) \\ 
\quad\quad\quad \left( {\lambda _{1,1,A}^{} x_k^{}  + \lambda _{1,1,B}^{} x_k^{}  + \lambda _{1,2,A}^{} x_{k - \tau _A }^{} } \right. \\ 
 \quad\quad\quad\left. { + \lambda _{1,2,B}^{} x_{k - \tau _B }^{}  + n_{k - \beta } } \right) ,\\ 
 \end{array} 
\end{equation} 
which yields
\begin{equation} \label{dec_var_long}
 \begin{array}{l}
D_{R,i}  =\sum\limits_{k = 1}^\beta  
 \Big(\left( {\lambda _{1,1,A}^2 s_{i,A}  + \lambda _{1,1,B}^2 s_{i,B} } \right)x_k^2  \\ 
\quad   + \left( {\lambda _{1,2,A}^2 s_{i,A} x_{k - \tau _A }^2  + \lambda _{1,2,B}^2 s_{i,B} x_{k - \tau _B }^2 } \right) \\ 
 \quad + \left( {s_{i,A}  + s_{i,B} } \right)\lambda _{1,1,A}^{} \lambda _{1,1,B}^{} x_k^2  \\ 
\quad + \left( {\lambda _{1,2,A}^{} x_{k - \tau _A }^{}  + \lambda _{1,2,B}^{} x_{k - \tau _B }^{} } \right)\lambda _{1,1,A}^{} s_{i,A} x_k^{}  \\ 
\quad + \left( {\lambda _{1,1,A}^{} x_k^{}  + \lambda _{1,1,B}^{} x_k^{} } \right)\lambda _{2,A}^{} s_{i,A} x_{k - \tau _A }^{}  \\ 
\quad+ \left( {s_{i,A}  + s_{i,B}  } \right) x_{k - \tau _A }^{} x_{k - \tau _B }^{}\lambda _{1,2,B}^{} \lambda _{1,2,A}^{}  \\ 
\quad+ \left( {\lambda _{1,2,A}^{} x_{k - \tau _A }^{}  + \lambda _{1,2,B}^{} x_{k - \tau _B }^{} } \right)\lambda _{1,1,B}^{} s_{i,B} x_k^{}  \\ 
\quad + \left( {\lambda _{1,1,A}^{} x_{k - \tau _A }^{}  + \lambda _{1,1,B}^{} x_{k - \tau _B }^{} } \right)\lambda _{1,2,B}^{} s_{i,B} x_{k - \tau _B }^{}  \\ 
\quad  + \left( {\lambda _{1,1,A}^{} s_{i,A} x_k^{}  + \lambda _{1,1,B}^{} s_{i,B} x_k^{} } \right)n_{k - \beta }  \\ 
\quad  + \left( {\lambda _{1,2,A}^{} s_{i,A} x_{k - \tau _A }^{}  + \lambda _{1,2,B}^{} s_{i,B} x_{k - \tau _B }^{} } \right)n_{k - \beta }  \\ 
\quad  + \left( {\lambda _{1,1,A}^{} x_k^{}  + \lambda _{1,1,B}^{} x_k^{} } \right)n_k  \\ 
\quad  + \left( {\lambda _{1,2,A}^{} x_{k - \tau _A }^{}  + \lambda _{1,2,B}^{} x_{k - \tau _B }^{} } \right)n_k  \\ 
\quad  + n_k n_{k - \beta } \Big)\cdot  \\ 
 \end{array}
 \end{equation}

The first two lines in equation \eqref{dec_var_long} represent the useful signal and the other components represent the intersymbol interference (ISI) and noise. It should be mentioned that the channel coefficients have zero mean and the mean of the decision variable is equal to the mean of the useful signal. Therefore, the mean of the decision variable can be represented as
\begin{equation} 
\label{mean_R}
\begin{array}{l}
E\left[ {D_{R,i} } \right] = \Big(  (\lambda _{1,1,A}^2 
+\lambda _{1,2,A}^2  ) s_{i,A} \\
 \quad  \quad   \quad   \quad  \,\,\, + (\lambda _{1,1,B}^2 + \lambda _{1,2,B}^2) s_{i,B}   \Big) E_b,
 \end{array}
\end{equation}
where $E_b  = 2\sum\limits_{k = 1}^\beta  {x_k^2 } $ is the transmitted bit energy. Note that for long spreading codes the bit energy $E_b$ might be assumed to be constant~\cite{Kad09ieee}. However, this assumption is not valid for short spreading codes due to the non-periodic and random behaviour of chaotic signals as reported in~\cite{Kad08cs}. Therefore, since the spreading factor in this work is sufficiently high, the constant bit energy assumption is valid.  
%
Note that the mean expression given in equation (\ref{mean_R}) is obtained based on the fact that chaotic signals, channel coefficients and AWGN are zero mean and the mean product of the delayed version of chaotic signals is also zero, i.e. $E\left[ {x_k x_{k - \tau } } \right] = 0$. However, the cross multiplication of the first user's reference signal with the data carrier signal of the second user results in a strong interference term given by the term $\left(  s_{i,A}  +  s_{i,B}  \right) \lambda _{1,1,A}\lambda _{1,1,B} E_b$ in equation (\ref{dec_var_long}). This term represents ISI, which can be strong enough to severely degrade the system performance.

\begin{table*}[htb!]
\small 
\caption{Network coding mapping scheme} 
\centering 
\begin{tabular}{|r| r| c| c| c| r| c| c| c |} 
\hline\hline 
 $e_{\mathcal{A}}\,\,\,\,\,\,\,$    & $e_{\mathcal{B}}\,\,\,\,\,\,\,\,$  &  $s_{\mathcal{R},A}\,\,\,\,\,\,\,\,$ &  $s_{\mathcal{R},B}$ &$s_{\mathcal{R},M}$ & $e_\mathcal{R}\,\,\,\,\,\,\,\,$ &  $s_{\mathcal{(A,B)},D}$  &   $\begin{array}{l} s_{\mathcal{(A,B)},M}=\\ \mid s_{\mathcal{(A,B)},D} +s_{B} \mid \end{array}$ &   $\begin{array}{l} \hat s_{\mathcal{A}}=\\s_{\mathcal{(A,B)},M}- 1 \end{array}$\\[0.05ex] \hline\hline
    $[x,1x]$    &   $[x,1x]$   & $1 $ &  1  & $\,\,1$ & $[x,1x]$ & $\,\,1$ &  2 & $1$ \\ \hline
    $[x,1x]$   &  $[x,-1x]$   & $1$ &  \hspace{-0.3 cm}$\,-1$ & \hspace{-0.2 cm}$\,-1$ & $[x,-1x]$ & \hspace{-0.2 cm}$\,-1$ &   2 &  $1$ \\ \hline
    $[x,-1x]$ &  $[x,1x]$    & \hspace{-0.3 cm}$\,-1$ &  1 & \hspace{-0.2 cm}$\,-1$ & $[x,-1x]$ & \hspace{-0.2 cm}$\,-1$ &  0 & \hspace{-0.3 cm}$\,-1$ \\ \hline
    $[x,-1x]$ &  $[x,-1x]$   & \hspace{-0.3 cm}$\,-1$&  \hspace{-0.3 cm}$\,-1$  & $\,\,1$ & $[x,1x]$ & $\,\,1$ & $ \,\,\,\, 0\,\,\,\,\,$ &$-1\,\,\,\,\,$  \\ \hline
  \end{tabular}\label{map-PNC_DCSK2}
\end{table*}

Additionally, there are $18$ other interference components in equation \eqref{dec_var_long} resulting from the cross multiplication of the users' signals. Therefore, considering all these interferences present in the decision variable, it can be inferred that in order to design reliable PNC-DCSK systems some form of interference mitigation is required.

\begin{figure}[htbp]
\centering 
\includegraphics[width=\linewidth]{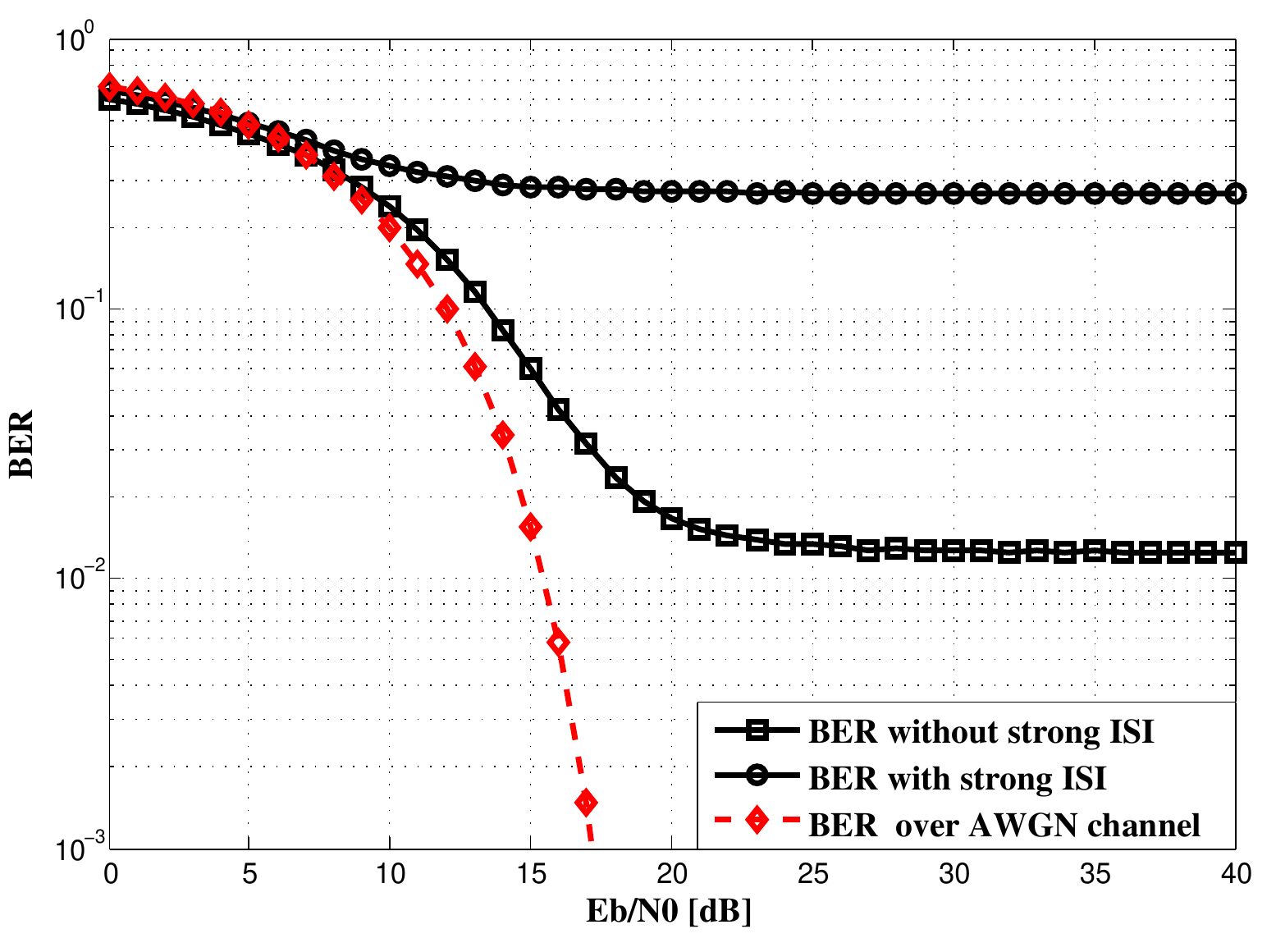}
\caption {BER performance at the relay of the PNC-DCSK scheme over multipath channel with different interference levels as well as over AWGN channel.
\label{BER_comparaison_PNC}} 
\end{figure}

Fig. \ref{BER_comparaison_PNC} depicts the BER performance at the relay of the PNC-DCSK scheme, when considering transmission over AWGN and multipath fading channels with different interference levels. In order to show the impact of a strong ISI on the system performance for multipath channel, the BER curves with and without the term $\left(  s_{i,A}  +  s_{i,B}  \right) \lambda _{1,1,A}\lambda _{1,1,B} E_b$ in equation (\ref{dec_var_long}) are plotted. This is done by extracting this term from the decision variable and assuming that it is known at the relay, which is impractical but has been done for the sake of elaboration. In Fig. \ref{BER_comparaison_PNC}, the BER results are obtained for a spreading factor $\beta=100$ and consider the following multipath channel parameters: $E[\lambda _{1,1,A}^{2}]= E[\lambda _{1,1,B}^{2}]= 0.9$, $E[\lambda _{1,2,A}^{2}]= E[\lambda _{1,2,B}^{2}]= 0.75$, $\tau _A=5$, and $\tau _B =8$ (i.e., $T_c=1$).

As shown in Fig. \ref{BER_comparaison_PNC}, the performance of the relay in the PNC-DCSK when considering transmission over AWGN channel is considerably better than the two other scenarios, where multipath channel is considered. This is mainly associated with the fact that for the AWGN channel the number of interference components are reduced significantly. On the other hand, as shown in Fig. \ref{BER_comparaison_PNC}, in multipath channels when strong ISI represented by the term $\left(  s_{i,A}  +  s_{i,B}  \right) \lambda _{1,1,A}\lambda _{1,1,B} E_b$ is taken into account in the decision variable, the performance of the relay is significantly degraded. Moreover, even if this strong ISI term is neglected, the other interference terms in equation (\ref{dec_var_long}) affect the relay performance. For example, for $E_b/N0 > 25$ dB the bit error rate is floored at $10^{-2}$ when we deliberately remove the strong ISI term. This poor performance result shows that the actual design with a DCSK modulator can not be considered as a potential application for PNC schemes. 

The above study indicates the limitation of the proposed scheme and the open problem to improve such a system performance by reducing its interference level and thus making the actual PNC-DCSK more reliable. Hence, in the following we propose potential solutions to design reliable PNC-DCSK techniques.

\section{Time and Frequency Multiplexed Network Coding Schemes for DCSK system}
The majority of the interference terms in equation (\ref{dec_var_long}) are generated from the cross product of the users' data signals and the reference signals. In what follows we show how the interference can be mitigated by using time domain or frequency domain multiplexing techniques to separate the signals transmitted by user $\mathcal{A}$ and $\mathcal{B}$ during the first phase of relaying. 

When time multiplexing is used during the first phase of relaying, the user nodes $\mathcal{A}$ and $\mathcal{B}$ transmit their signals to the relay $\mathcal{R}$ using different time slots, $T_A$ and $T_B$. We refer to this scheme as scheme $2$. Hence, in order to avoid any signal superposition, the two time slots must satisfy the following relation $\vert T_A - T_B \vert > 2\beta T_c $. The time slot is defined as the required time for a source or relay to transmit one DCSK symbol. In this scheme the user synchronization is not required and the required number of time slots to achieve the end-to-end communication is $3$.

On the other hand in scheme $3$ where the signals superposition is carried out in frequency domain, the two users transmit their signals to the relay at the same time but using different carrier frequencies, $f_A$ and $f_B$. Note that when frequency multiplexing is used, the users' frequencies are required to meet the specification of $\vert f_A - f_B \vert > W_s$, where $W_s \approx 1/T_c$ denotes the signal bandwidth. Similarly to scheme $2$, scheme $3$ does not require any users synchronization process and the total time slots required to achieve the end-to-end transmission is equal to $2$. Additionally, it is essential to note that scheme $3$ requires twice the bandwidth needed for scheme $1$ or scheme $2$.

\begin{table*}[htb!]
\small 
\caption{Complexity parameters of the proposed network coding scheme} 
\centering 
\begin{tabular}{|l| c| c| c| c| c| c| c| c |} 
\hline\hline 
 Schemes\,\,\,\,\,\,    &  Decoding times\,\,\,  &  Mapping times\,\, & Modulation times\, & Users sync. & BW &  Time slots  \\\hline\hline
    scheme $1$   &   $P$   &   $P$  &    $P$       &  Yes  & $W_s$ & $2$  \\ \hline
    scheme $2$   &  $2P$   &   $P$  &    $P$       &  No    & $W_s$ & $3$   \\ \hline
    scheme $3$   &  $2P$   &   $P$  &    $P$       &  No    & $2W_s$ & $2$    \\ \hline
 \hline
   \end{tabular}\label{complx}
\end{table*}

The use of time or frequency multiplexing results in eliminating the cross product between the users' signals, which contributes to the mitigation of interference. In both schemes, the relay decodes and maps the received signals separately, before broadcasting the resultant DCSK frame, which leads to a considerably lower interference than in scheme $1$. 

Since the access protocol is the same for time and frequency multiplexing schemes, Table \ref{map-PNC_DCSK2} summarizes the decoding and mapping process that is valid for both scheme $2$ and scheme $3$. According to this table, in both scenarios, the relay separately decodes the information sent by the two nodes given by $s_{\mathcal{R},A}$ and $s_{\mathcal{R},B}$. After the bits are  mapped into a bit steam $s_{\mathcal{R},M}$, the relay modulates the mapped bits into a DCSK signal frame $e_{\mathcal{R}}$ and transmits it to the two users in the second phase of relaying. For instance, in order to recover the transmitted data sent by user $\mathcal{A}$ to user $\mathcal{B}$ side, which is represented by $\hat s_{A}$ in Table~\ref{map-PNC_DCSK2}, the received signal from the relay is first despread and decoded by user $\mathcal{B}$. The decoded symbols $s_{\mathcal{(A,B)},D}$ are then demapped into $s_{\mathcal{(A,B)},M}$ by using the same mapping function as the relay. Finally, user $\mathcal{B}$ can recover the signal transmitted by user $\mathcal{A}$ by subtracting its own data signal. 

\subsection{Complexity analysis}
In this section we summarize the complexity analysis of schemes $1$, $2$, and $3$. The decoding complexity in the three schemes is identical and it is equivalent to that of the conventional spread spectrum communication system. Hence, the complexity of the different schemes is evaluated by the number of decoding, mapping and modulation operations performed at the relay.

Let us assume that each user transmits a packet of $P$ bits during the first phase of relaying. Hence, since the signals are superimposed in the wireless channel, the relay in scheme $1$ needs $P$ decoding operations, $P$ mapping operations and $P$ modulation operations. However, in schemes $2$ and $3$, the relay decodes the user signals separately after the first transmission phase. Hence, the required decoding, mapping and modulating operations required for these two schemes are $2P$, $P$, and $P$, respectively. 

Additionally, it should be noted that the user synchronization is essential for scheme $1$, while it is not needed for schemes $2$ and $3$. Table \ref{complx} compares the complexity, the systems parameters of the proposed schemes and the number of required time slots to achieve an end-to-end transmission. 

Apart from synchronization, it can be concluded that while schemes $2$ and $3$ have roughly the same complexity to achieve an end-to-end transmission, scheme $3$ requires twice the bandwidth and less time slots as compared to scheme $2$. Therefore, the choice between the two schemes is mainly based on the user requirements particularly in terms of time or bandwidth.

\subsection{\textbf{BER performance analysis of the frequency and time multiplexed network coding schemes for DCSK system}}
\label{sec:perf}
In this section, the analytical BER end-to-end expression is derived for schemes $2$ and $3$. In both schemes the signals are separately decoded by the relay during the first phase of relaying, hence the BER derivation methodology for a given wireless channel is the same for the two schemes.

We Consider the scenario where user $\mathcal{A}$ transmits its data to user $\mathcal{B}$ via the relay $\mathcal{R}$. We will derive the BER performanc of user $\mathcal{B}$. In this case, user $\mathcal{B}$ first despreads and decodes the received mapped symbols sent by the relay and then extracts its own data to recover the useful data transmitted by user $\mathcal{A}$, as shown in Table \ref{map-PNC_DCSK2}. 
In our analysis we assume equal transmit power for all nodes. In this sense, $\rm BER_{1,A}$, and $ \rm BER_{1,B}$ represent the bit error rates of the users $\mathcal{A}$ and $\mathcal{B}$ in the first transmission phase, respectively, and $ \rm BER_{2,B}$ denotes the bit error rate of user $\mathcal{B}$ in the second transmission phase.

The bit error rate at the relay in the first transmission phase is the sum of two error functions.   The first error function represents the  situation in which there is  no bit error detection from the user $\mathcal{B}$, but an error  in the detection of the user $\mathcal{A}$'s bit. Oppositely, the second situation represented in equation \eqref{ber_relay} by the second error function is  when there is no error detection in the bit transmitted by user $\mathcal{A}$ to the relay, but there is an error in the transmitted bit by user $\mathcal{B}$. Hence, the bit error rate at the relay can be represented as:
\begin{equation} 
\label{ber_relay}
\rm BER_{1,R}\hspace{-0.05 cm}= \hspace{-0.05 cm}\rm BER_{1,A}(1\hspace{-0.02 cm}-\rm BER_{1,B})\hspace{-0.01 cm} +\hspace{-0.01 cm}\rm BER_{1,B}(1\hspace{-0.02 cm}-\rm BER_{1,A})\cdot
\end{equation}

Additionally, the end-to-end BER can be determined as sum of two possible scenarios as follow:
\begin{equation} 
\label{ber_e2e1}
 \rm BER= \rm BER_{1,R}(1-\rm BER_{2,B}) + \rm BER_{2,B}(1- \rm BER_{1,R}),
\end{equation}
which yields the final end-to-end BER expression of
\begin{equation} 
\label{ber_e2e2}
\begin{array}{l}
\hspace{-0.1 cm}\rm BER =  \rm BER_{1,A}+ \rm BER_{1,B} + \rm BER_{2,B} \\
  \quad\quad -2 \rm BER_{1,A} \rm BER_{1,B} -2 \rm BER_{1,A} \rm BER_{2,B} \\
\quad\quad -2 \rm BER_{1,B} \rm BER_{2,B}\hspace{-0.06 cm} + \hspace{-0.05 cm}4 \rm BER_{1,A} \rm BER_{1,B} \rm BER_{2,B}\,\cdot
\end{array}
\end{equation}

Equation \eqref{ber_e2e2} requires the computation of three different BER expressions, $BER_{1,A}$, $BER_{1,B}$, and $BER_{2,B}$. To compute the end-to-end bit error rate expression for the first transmission phase of user $A$, the statistical properties of the appropriate decision variable must be determined. In this case, the decision variable for the $i^{th}$ bit transmitted by user $\mathcal{A}$ in the first phase might be given by
\begin{equation} 
\label{dec_relay}
\begin{array}{l}
 D_{i,1,A}  =  \left( \lambda _{1, 1,A}^{} s_{i,A} x_k^{}   + \lambda _{1,2,A}^{} s_{i,A} x_{k - \tau _A }^{}    + n_k  \right) \\ 
\quad\quad \quad\,\,\,\,\,\,\left( \lambda _{1, 1,A}^{} x_k^{}   + \lambda _{1, 2,A}^{} x_{k - \tau _A }^{}   + n_{k - \beta }  \right). 
 \end{array}
\end{equation} 

Therefore, the decision variable can be formulated as
\begin{equation} 
\label{dec_a}
\begin{array}{l}
\hspace{-0.4 cm} D_{i,1,A}  = \sum\limits_{k = 1}^\beta   \Big( \lambda _{1, 1,A} x_k^2 s_{i,A}  + \lambda _{1,2,A} x_{k - \tau _A }^2 s_{i,A}  \\ 
 \quad\quad\,\,\, + 2\lambda _{1,A} \lambda _{2,A} x_k^{} x_{k - \tau _A }^{} s_{i,A}  \\ 
 \quad\quad\,\,\,  + \left( {\lambda _{1,1,A} x_k^{} s_{i,A}  + \lambda _{1,2,A} x_{k - \tau _A }^{} s_{i,A} } \right)n_{k - \beta }  \\ 
  \quad\quad\,\,\, + \left( {\lambda _{1,1,A} x_k^{}  + \lambda _{1,2,A} x_{k - \tau _A }^{} } \right)n_k + n_k n_{k - \beta } \Big) ,\\ 
 \end{array} 
\end{equation}
where the first line denotes the useful signal and the other lines represent the multipath and the Gaussian noise due to the interfering signals. 

The decision variable in this scenario follows a Gaussian distribution, because the sum of the product of two chaotic sequences is zero which eliminates the contribution of the second line in equation~\eqref{dec_a}, i.e. $\sum\limits_{k = 1}^\beta \lambda _{1,A} \lambda _{2,A} x_k^{} x_{k - \tau _A }^{} s_{i,A} \approx 0$ and all the other interference components follow the Gaussian distribution. Additionally, the channels coefficients are considered to be independent and the spreading sequences and the Gaussian noises are also independent. Furthermore, the noise samples are uncorrelated and independent of the chaotic sequences and the chaotic samples themselves are independent from each other. Therefore, it can be concluded that the terms corresponding to different signal interferences present in the decision variable expression given by equation \eqref{dec_a} are independent.

Knowing that the chaotic and noise signals are zero mean, the mean of the decision variable for a given $i^{th}$ bit can be defined as
\begin{equation} 
\label{mean_v1}
E\left[ {  D_{i,1,A} } \right] =   \frac{ (\lambda _{1,1,A}^2 +\lambda _{1,2,A}^2  )s_{i,A} E_b}{2},
\end{equation}
and the total variance of the decision variable is the sum of the different interference components.  
\begin{equation}
\label{Equ:Var_tot}
V\left[ { D_{i,1,A}   } \right] = V\left[ C \right] + V\left[ D \right] + V\left[ E \right] + V[F]
\end{equation}
where
\[
C=  \sum\limits_{k = 1}^\beta  2\lambda _{1,1,A} \lambda _{1,2,A} x_k^{} x_{k - \tau _A }^{} s_{i,A}, 
\]
\[
D= \sum\limits_{k = 1}^\beta  \left( {\lambda _{1,1,A} x_k^{} s_{i,A}  + \lambda _{1,2,A} x_{k - \tau _A }^{} s_{i,A}, } \right)n_{k - \beta },
\]
\[
E= \sum\limits_{k = 1}^\beta  \left( {\lambda _{1,1,A} x_k^{}  + \lambda _{1,2,A} x_{k - \tau _A }^{} } \right)n_k, 
\]
and
\[
F= \sum\limits_{k = 1}^\beta n_k n_{k - \beta }.
\]

Based on the independent and uncorrelated characteristics mentioned earlier, the variance terms in equation \eqref{Equ:Var_tot} might be formulated as
\begin{equation} \label{eq_c}
V\left[ C \right] = 2\lambda _{1,1,A}^{2}\lambda _{1,2,A}^{2} E_b ,
\end{equation}

Equation \eqref{eq_c} is obtained because the considered chaotic map has a normalized variance $\textrm{E}[x^2]=1$ which makes  the expression $E\left[ \sum\limits_{k = 1}^\beta ( x_k^{} x_{k - \tau _A })^2\right]$ equivalent to $E\left[ \sum\limits_{k = 1}^\beta ( x_k^{})^2\right]$  and equal to $E_b/2$ 

\begin{equation}
V\left[ D \right] = \frac{(\lambda _{1,1,A}^{2} + \lambda _{1,2,A}^{2})N_0 E_b}{4} ,
\end{equation}
\begin{equation}
V\left[ E \right] = \frac{(\lambda _{1,1,A}^{2} + \lambda _{1,2,A}^{2})N_0 E_b}{4} ,
\end{equation}
and
\begin{equation}
V\left[ F \right] = \frac{{ \beta N_0^2 }}{4} \cdot
\end{equation}

Hence, the total variance of the decision variable given by equation \eqref{Equ:Var_tot} can be formulated as
\begin{equation} 
\label{var}
\begin{array}{l}
V\left[ { D_{i,1,A}  } \right] =  2\lambda _{1,1,A}^{2}\lambda _{1,2,A}^{2} E_b \\
\quad\quad\quad\quad\quad\,+ \frac{(\lambda _{1,1,A}^{2} + \lambda _{1,2,A}^{2})N_0 E_b}{2}+  \frac{{\beta N_0^2 }}{4}\cdot
\end{array}
\end{equation}

Therefore, the $\rm BER_{1,A}$ for user $\mathcal{A}$ during the first transmission phase can be obtained by
\begin{equation}\label{ber_ga_ex1}
\begin{array}{l}
\rm BER_{1,A}  = \,\,\frac{1}{2}\Pr \left( {\left. {\textit{D}_{i,1,A}  < 0} \right|s_{i,A}  =  + 1} \right)\vspace{0.1 cm} \\
\quad\quad\quad\quad\,\, + \frac{1}{2}\Pr \left( {\left. {\textit{D}_{i,1,A}  > 0} \right|\rm s_{i,A}  = \hspace{-0.01 cm}- 1} \right) , 
 \end{array}
\end{equation}
which results in
\begin{equation}\label{ber_ga_ex}
\rm BER_{1,A}= \frac{1}{2} \rm erfc\left( {\frac{{E\left[ {\left. {\textit{D}_{i,1,A} } \right|s_{i,A}  =  + 1} \right]}}{{\sqrt {2{\mathop{\rm Var}} \left[ {\left. {\textit{D}_{i,1,A} } \right|s_{i,A}  =  + 1} \right]} }}} \right),
\end{equation}
where $ \rm erfc(x)$ is the complementary error function defined by 
\begin{equation}
\rm erfc(x) \equiv \frac{2}{{\sqrt \pi  }}\int_x^\infty  {e^{ - \mu ^2 } } d\mu \cdot
\end{equation}
Therefore, the instantaneous BER expression of user $\mathcal{A}$ is given by   
\begin{equation} \label{BER_1}
\begin{array}{l}
\hspace{-0.2 cm}\rm BER_{1,A} (\lambda _{1,1,A}, \lambda _{1,2,A}) = \frac{1}{2} \rm erfc\left( \Big[ \frac{{16\lambda _{1,1,A}^2 \lambda _{1,2,A}^2 E_b }}{{\left( {\lambda _{1,1,A}^2  + \lambda _{1,2,A}^2 } \right)^2 E_b^2 }}\right. \vspace{0.1 cm}\\
\,\,\,\left.+ \frac{{4 \left( {\lambda _{1,1,A}^2  + \lambda _{1,2,A}^2 } \right)N_0 E_b^{} }}{{\left( {\lambda _{1,1,A}^2  + \lambda _{1,2,A}^2 } \right)^2 E_b^2 }} + \frac{{2\beta N_0^2 }}{{\left( {\lambda _{1,1,A}^2  + \lambda _{1,2,A}^2 } \right)^2 E_b^2 }} \Big]^{-\frac{{1}}{2}} \right) ,
 \end{array}
\end{equation}
which can be simplified to 
\begin{equation} \label{BER_2}
\begin{array}{l}
\hspace{-0.2 cm}\rm BER_{1,A} (\lambda _{1,1,A}, \lambda _{1,2,A})  = \frac{1}{2} \rm erfc\left( \Big[ \frac{{16\lambda _{1,1,A}^2 \lambda _{1,2,A}^2 }}{{\left( {\lambda _{1,1,A}^2  + \lambda _{1,2,A}^2 } \right)^2 E_b }}\right. \vspace{0.1 cm}\\
\quad\,\,\,\,\left. + \frac{{4N_0 }}{{\left( {\lambda _{1,1,A}^2  + \lambda _{1,2,A}^2 } \right)E_b^{} }} + \frac{{2\beta N_0^2 }}{{\left( {\lambda _{1,1,A}^2  + \lambda _{1,2,A}^2 } \right)^2 E_b^2 }} \Big]^{-\frac{{  1}}{2}}  \right) \cdot
 \end{array}
\end{equation}

If the largest multipath time delay is shorter than the bit duration $0 <\tau_{1,A} << \beta T_c$, the ISI can be neglected compared to the interference within each symbol due to multipath delay. However, if the delay $\tau_{A}$ increases, the ISI increases significantly. Hence, for large values of $\tau_{A}$ the hypothesis of neglecting the ISI is not valid. The condition $0 <\tau_{1,A} << \beta T_c$ is commonly used in the literature~\cite{Xia04,Chen13,kad13arx}, where the ISI can be neglected under this condition. 
Similarly, these studies have shown that for a large spreading factor we have
\begin{equation} \label{eq_ss}
\sum\limits_{k = 1}^\beta  {\left( {x_{k - \tau _l } x_{k - \tau _j } } \right)}  \approx 0 \:\:\: for \: l \ne j \cdot
\end{equation} 

In our work the variable $C$ representing the ISI can be neglected, since $C\approx 0$. Therefore, the BER expression in equation \eqref{BER_2} may be expressed as
\begin{equation}
\begin{array}{l}
\rm  BER_{1,A} (\gamma _{1,A}) =   
 \frac{1}{2} \rm erfc\left( {\left[ {\frac{4}{{\gamma _{1,A}^{} }} + \frac{2\beta }{{\gamma _{1,A}^2 }}} \right]^{-\frac{{  1}}{2}} } \right), 
 \end{array}
\end{equation}
where $\gamma _{1,A}^{}  = \frac{{\left( {\lambda _{1,1,A}^2  + \lambda _{1,2,A}^2 } \right)E_b^{} }}{{N_0 }} $ represents the signal-to-noise ratio (SNR) at the relay side. Let ${\bar \gamma _{1,1} }$ and ${\bar \gamma _{1,2} }$ denote the average SNR of the received signal at the relay, then $\bar \gamma _{1,1}  = \frac{{E_b }}{{N_0 }}{\rm E}[\lambda _{1,1,A} ^2  ]$ and $\bar \gamma _{1,2}  = \frac{{E_b }}{{N_0 }}{\rm E}[\lambda _{1,2,A} ^2 ]$. 

For non-identical channel coefficients, i.e., $\bar \gamma _{1,1}  \ne \bar \gamma _{1,2}$, $f\left( {\gamma _{1,A} } \right)$ can be obtained by
\begin{equation}
f\left( {\gamma _{1,A} } \right) = \frac{1}{{\bar \gamma _{1,1} \hspace{-0.05 cm} - \hspace{-0.05 cm} \bar \gamma _{1,2} }}\hspace{-0.05 cm}\left( \hspace{-0.05 cm}{{\rm{exp}}\big( { - \frac{{\gamma_{1,A} }}{{\bar \gamma _{1,1} }}} \big){\rm{ - exp}}\big( { - \frac{{\gamma _{1,A} }}{{\bar \gamma _{1,2} }}} \big)} \hspace{-0.05 cm}\right).
\end{equation}
On the other hand, for identical channels, i.e., $\bar \gamma _{1,1} = \bar \gamma _{1,2}$, $f\left( {\gamma _{1,A} } \right)$ is given by
\begin{equation}
f\left( {\gamma _{1,A} } \right) =  \frac{{\gamma _{1,A} }}{{\bar \gamma _{1,1} ^2 }}{\rm{exp}}\left( { - \frac{{\gamma_{1,A} }}{{\bar \gamma _{1,1} }}} \right)\cdot
\end{equation}
Eventually, the $\rm BER_{1,A}$ can be obtained by averaging the conditional $ \rm BER_{1,A}$  as follows
\begin{equation} \label{ber1a}
\rm BER_{1,A}  = \int_0^\infty   \rm {BER_{1,A} \left( {\gamma_{1,A} } \right)} f\left( {\gamma _{1,A} } \right)d\gamma _{1,A} .
\end{equation}

The BER expression for user $\mathcal{B}$ in the first transmission phase can be derived in a similar manner to that of user $\mathcal{A}$, when considering communications over a multipath channel with different channel coefficients and different delay values for each user. Hence, the instantaneous BER for user $\mathcal{B}$ can be expressed as
\begin{equation}
\begin{array}{l}
\rm  BER_{1,B} (\gamma _{1,B}) =   \frac{1}{2} \rm erfc\left( {\left[ {\frac{4}{{\gamma _{1,B}^{} }} + \frac{2\beta }{{\gamma _{1,B}^2 }}} \right]^{-\frac{{ 1}}{2}} } \right),  
 \end{array}
\end{equation}
where $\gamma _{1,B}^{}  = \frac{{\left( {\lambda _{1,1,B}^2  + \lambda _{1,2,B}^2 } \right)E_b^{} }}{{N_0 }}$ represents the SNR at the relay.
Finally, similar to user $\mathcal{A}$, the final BER expression for user $\mathcal{B}$ in the first transmission phase is obtained by averaging the conditional bit error rate function as follows
\begin{equation} \label{ber1b}
\rm BER_{1,B}  = \int_0^\infty  {BER_{1,B} \left( {\gamma_{1,B} } \right)} f\left( {\gamma _{1,B} } \right)d\gamma _{1,B}.
\end{equation}

Similarly, the instantaneous BER expression for user $\mathcal{B}$ in the second transmission phase can be expressed as
\begin{equation}
\begin{array}{l}
 \rm BER_{2,B} (\gamma _{2,A}) = \frac{1}{2} \rm erfc\left( {\left[ {\frac{4}{{\gamma _{2,B}^{} }} + \frac{2\beta }{{\gamma _{2,B}^2 }}} \right]^{-\frac{{  1}}{2}} } \right), 
 \end{array}
\end{equation}
where $\gamma _{2,B}^{}  = \frac{{\left( {\lambda _{2,1,B}^2  + \lambda _{2,2,A}^2 } \right)E_b^{} }}{{N_0 }} $ represents the SNR at user $\mathcal{B}$. Hence, the final $\rm BER_{2,B}$ expression might be formulated as
\begin{equation} \label{ber2b}
\rm BER_{2,B}  = \int_0^\infty  { \rm BER_{2,B} \left( {\gamma_{2,B} } \right)} f\left( {\gamma _{2,B} } \right)d\gamma _{2,B}.
\end{equation}

Finally, the analytical end-to-end BER expression of the proposed schemes $2$ and $3$ over multipath channel can be obtained by substituting equations \eqref{ber1a}, \eqref{ber1b}, and \eqref{ber2b} in equation \eqref{ber_e2e2}. 

\subsection{\textbf{Special case: One user is in a low interference zone and the other user is in high interference zone }}
In this section we analyse the network performance for two special scenarios: the first is when user $\mathcal{A}$ is in low interference zone while user $\mathcal{B}$ is in high interference zone and the second scenario is when user $\mathcal{B}$ is in low interference zone while user $\mathcal{A}$ is in high interference zone.

In the first scenario, when the user $\mathcal{A}$ is in low interference zone and user $\mathcal{B}$ is in high one, user $\mathcal{A}$ is affected by AWGN only, while user $\mathcal{B}$'s signal is transmitted over a multipath channel in addition to the AWGN. Therefore, in this case the channel coefficients in the $\rm BER_{1,A}$ expression have the values of  $\lambda_{1,1,A}=1$ and $\lambda_{1,2,A}=0$. Hence, the $\rm BER_{1,A}$ expression of equation \eqref{BER_2} can be simplified as
\begin{equation} \label{BER_a_awgn}
\rm BER_{1,A}   = \frac{1}{2} \rm erfc\left( \Big[ \frac{{4N_0 }}{{E_b^{} }} + \frac{{2\beta N_0^2 }}{{E_b^2 }} \Big]^{-\frac{{ 1}}{2}}  \right).
\end{equation}
And then we can substitute equation \eqref{BER_a_awgn} in equation \eqref{ber_relay} in order to get the end-to-end BER expression given in equation \eqref{ber_e2e2}.

On the other hand, in the second scenario, when user $\mathcal{B}$ is in low interference zone, equation \eqref{ber1a} remains the same and the two BER expressions of equations \eqref{ber1b} and \eqref{ber2b} can be expressed as
 \begin{equation} \label{BER_b_awgn}
\rm BER_{1,B} =BER_{2,B}  = \frac{1}{2} \rm erfc\left( \Big[ \frac{{4N_0 }}{{E_b^{} }} + \frac{{2\beta N_0^2 }}{{E_b^2 }} \Big]^{-\frac{{  1}}{2}}  \right) \cdot
\end{equation}
%

\subsection{\textbf{Throughput and link spectral efficiency analysis}}
We have shown in Section~\ref{sec:perf} that the end-to-end BER performance is the same for scheme $2$ and $3$, when they experience the same channel profiles. In this section, we analyse and compare the throughput and link spectral efficiency of the two schemes. Typically, the effective throughput $R_t$ for a constellation size $M$ is defined as the number of correct bits that a user receives per unit of time~\cite{Tse05}, which can be expressed as
\begin{equation} \label{Rt}
R_t = \frac{{ \rm log_2(M)(1 - \rm BER)}}{{T_n }},
\end{equation}
where $T_n$ is the time to exchange bits between the nodes, $(1-\rm BER)$ denotes the correct bits ratio received within a period of time $T_n$ and $M$ is the constellation size which is equal to $2$ for DCSK modulation.

For the frequency multiplexing DCSK scheme $3$, two time slots ($T_n=2T_b$) are required to exchange data between users. Hence, the total required time $T_n$ for transmission of a DCSK symbol can be formulated as
\begin{equation}
T_n= 4\beta T_c.
\end{equation}
In contrast, exchanging data between users in the time multiplexed DCSK scheme $2$ requires three time slots ($T_n=3T_b$), which results in the $T_n$ being formulated as
\begin{equation}
T_n= 6\beta T_c\cdot
\end{equation}

Therefore, in case of equal BER performance, the throughput in the frequency multiplexing DCSK scheme $3$ is higher than that for the time multiplexing DCSK scheme $2$. However, the bandwidth required for the two schemes is not the same and hence it is essential to analyse and compare the the link spectral efficiency of the two schemes. The link spectral efficiency $\Gamma$ is defined as the ratio of the maximum throughput of the system and the used bandwidth of the data link and can be formulated as
\begin{equation}
\Gamma = \frac{R_t}{W},
\end{equation}
where $\Gamma$ is the link spectral efficiency and $W$ is the total user bandwidth.  

For the time domain multiplexing scheme $2$, the total bandwidth required is equal to the DCSK bandwidth, which results in
\begin{equation}
W_s\approx\frac{1}{T_c},
\end{equation}
where $T_c$ is the time chip.

On the other hand, the total bandwidth $W_s^{'}$ for the frequency domain multiplexing scheme is twice the DCSK bandwidth, i.e. $3$ $W_s^{'}=2W_s$ and hence
\begin{equation}
W_s^{'}\approx \frac{2}{T_c}\cdot
\end{equation}

Consequently, the link spectral efficiency of frequency and time multiplexed network coding schemes for a DCSK system can be obtained as
\begin{equation}  
\label{link_f}
\Gamma_f = \frac{(1- \rm BER)}{8\beta},
\end{equation}
and
\begin{equation} \label{link_t}
\Gamma_t = \frac{(1-\rm BER)}{6\beta}, 
\end{equation}
where $\Gamma_f$ and $\Gamma_t$ are the link spectral efficiency for scheme $3$ and scheme $2$, respectively.

Therefore, according to equations \eqref{Rt}, \eqref{link_f} and \eqref{link_t}, considering performance at the same bit error rate, the throughput of the frequency multiplexed scheme $3$ is $1.5$ times higher than that of the time domain multiplexed scheme $2$. However, the link spectral efficiency of scheme $3$ is $0.75$ of that of the time domain scheme $2$. Therefore, the system designer would have to choose between scheme $2$ and scheme $3$ according to their requirements.

\section{Performance Results}
%
\begin{table*}[htb!]
\small 
\caption{Simulation parameters  } 
\centering 
\begin{tabular}{|l| c| c|  c| c| c| c| c| c| c| c| c |} 
\hline\hline 
$\beta$ & $E[\lambda_{1,1,A}^2]$ & $E[\lambda_{1,2,A}^2]$ & $\tau_{A}$ & $E[\lambda_{1,1,B}^2]$ &   $E[\lambda_{1,2,B}^2]$ & $\tau_{B}$ & $E[\lambda_{2,1,B}^2]$ & $E[\lambda_{2,2,B}^2]$ & $\tau_{B^{'}}$\\ 
\hline
25 & 0.7 & 0.89  & 3  &0.82  & 0.4  & 8  & 0.83  & 0.35  & 5\\
\hline

50 &  0.77  &0.47  & 4  &0.57  & 0.37  & 6  & 0.8  & 0.5  & 9\\
\hline

150  & 0.54  & 0.26  & 11  & 0.74  & 0.43  & 13  & 0.6  & 0.3  & 7\\
\hline
\end{tabular}
\label{param}
\end{table*}

In this section we will present the performance of the different proposed schemes for different system parameters, while comparing the proposed theoretical BER analysis with simulation results. Table \ref{param} lists the channel parameters used, where $E[\lambda_{1,1,A}^2]$, $E[\lambda_{1,2,A}^2]$, $E[\lambda_{1,1,B}^2]$, $E[\lambda_{1,2,B}^2]$, $E[\lambda_{2,1,B}^2]$ and $E[\lambda_{2,2,B}^2]$ denote the average power gains for different paths of the channels corresponding to user $\mathcal{A}$ and $\mathcal{B}$ during the fist and the second transmission phases. Additionally, $\tau_{A}$ and $\tau_{B}$ represent the delay spread during the fist transmission phase, while $\tau_{B^{'}}$ is the delay spread in the second phase of relaying. 

\begin{figure}[htbp]
\centering 
\includegraphics[width=\linewidth]{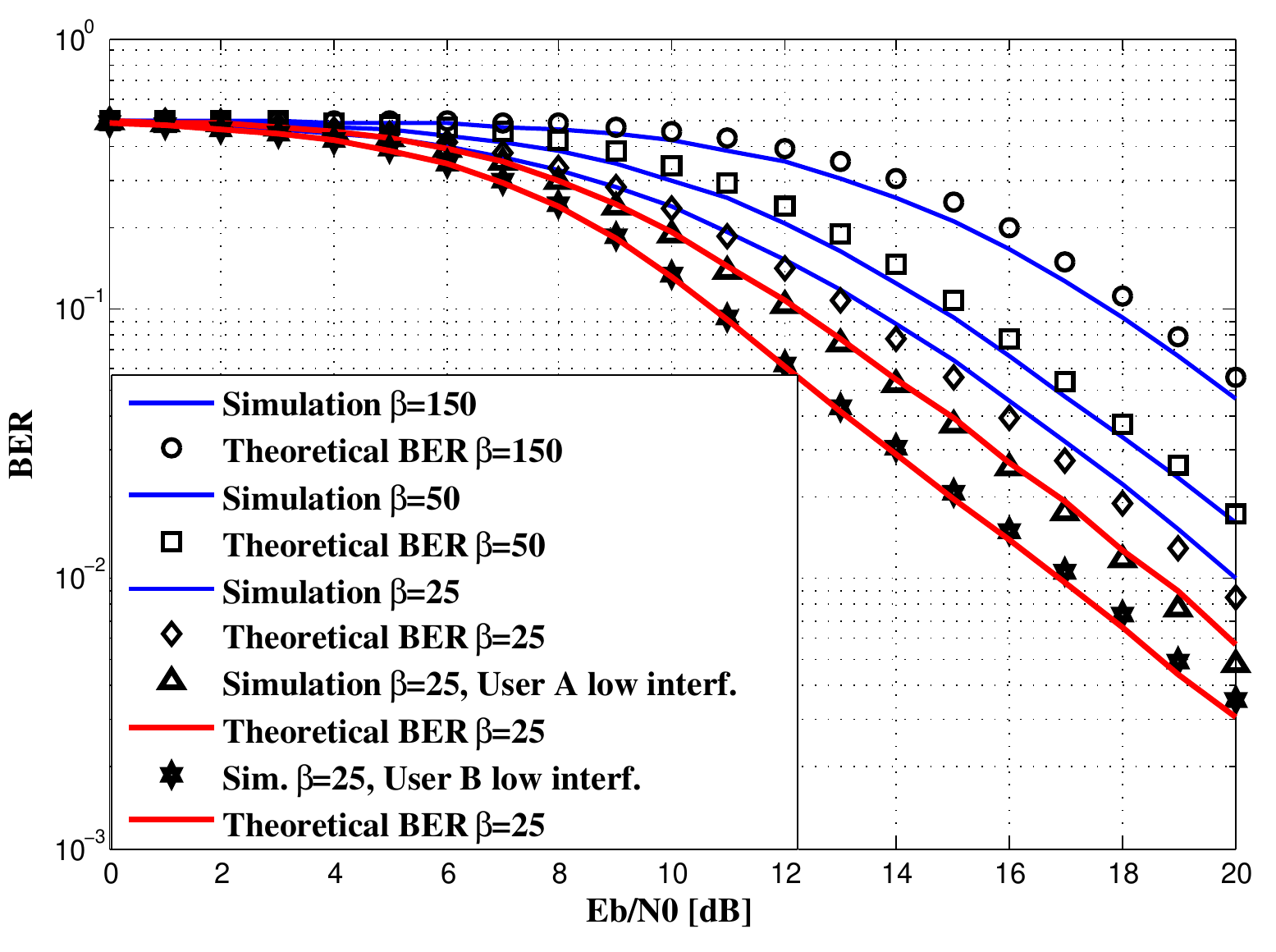}
\caption {Comparison of the Monte Carlo simulations and the theoretical end-to-end BER of schemes $2$ and $3$, while considering transmission over multipath Rayleigh fading channel for different spreading factors, channel gains and delay spreads.  
\label{ber_25_50_150}} 
\end{figure}

Fig. \ref{ber_25_50_150} shows the Monte Carlo simulations and the theoretical end-to-end BER performance for Schemes $2$ and $3$, while communication over multipath Rayleigh fading channel using the system parameters summarized in table \ref{param}. As shown in the figure, the simulation results close follow the analytical BER expressions for all spreading factors, channel gains and delay spreads. The close match between the simulation results and the analytical curves validates the assumption of neglecting the ISI. Moreover, Fig. \ref{ber_25_50_150} shows the performance of the two specific scenarios studied in subsection IV-C, namely when one user has low interference level while the other user has high interference. The lower bound performance of this network is obtained when the user $\mathcal{A}$ transmits its data to user $\mathcal{B}$, while user $\mathcal{B}$ is in low interference zone or when the user $\mathcal{B}$ transmits its data to user $\mathcal{A}$, while user $\mathcal{A}$ is in low interference zone.

\begin{figure}[htbp]
\centering 
\includegraphics[width=\linewidth]{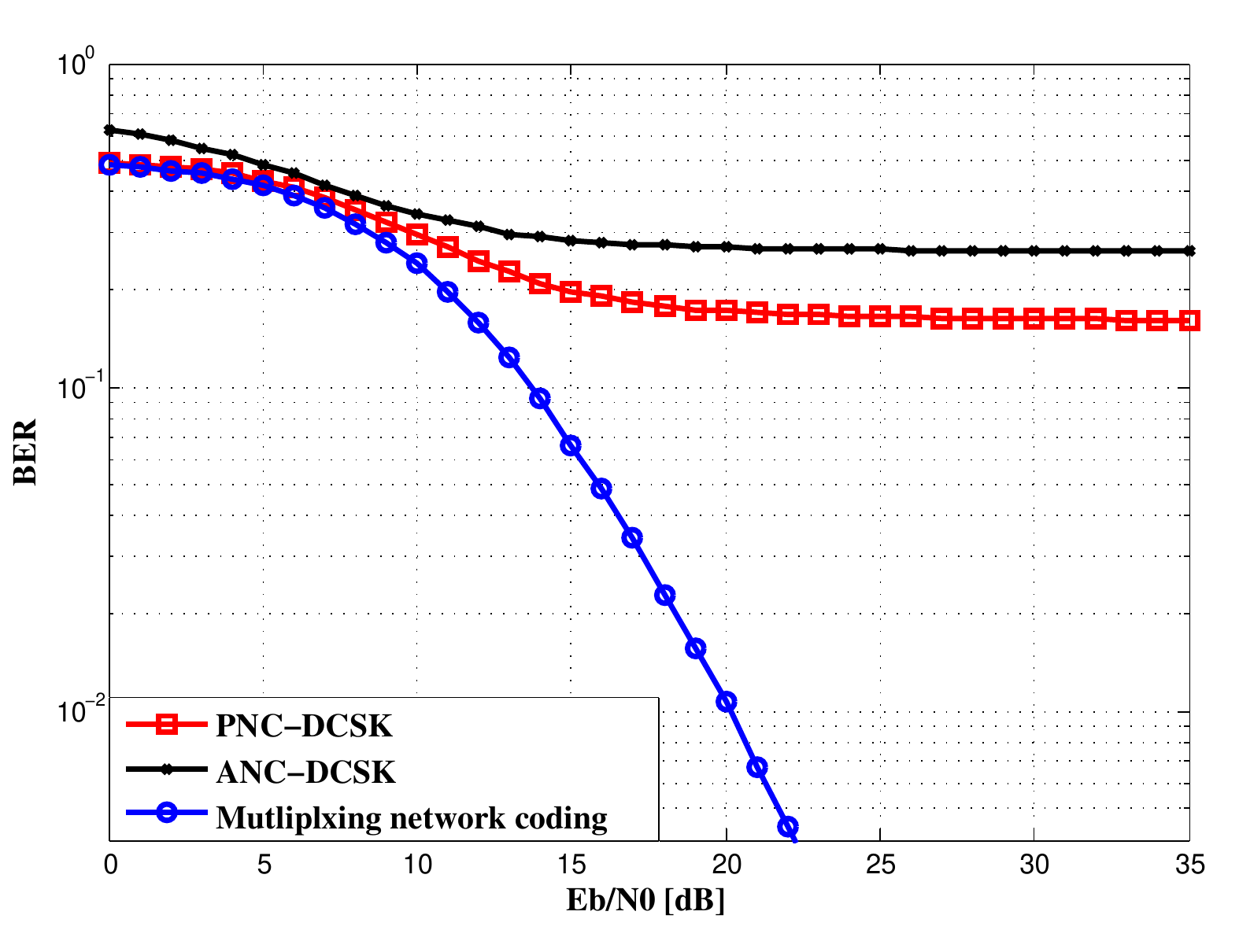}
\caption {BER performance comparison between the PNC-DCSK scheme $1$ and the multiplexed network coding schemes $2$ and $3$.
\label{BER__PNC_mux}} 
\end{figure} 

In this paper, we have not focused on the optimal spreading factor length for DCSK system due to the fact that this has been the subject of many research work conducted on the performance of DCSK systems for different spreading factors. For instance, the optimal spreading factor for a single carrier DCSK system was studied in \cite{Lau03} and for multi-carrier DCSK in \cite{6560492}.

Fig. \ref{BER__PNC_mux} compares the simulated BER performance of the end-to-end BER for ANC-DCSK, PNC-DCSK system and the proposed schemes $2$ and $3$, when communicating over multipath Rayleigh fading channel. In these results, the spreading factor used is $\beta=25$ in addition to the channel parameters listed in the table \ref{param}. The figure shows that the multiplexed network coding schemes $2$ and $3$ outperform the PNC-DCSK and the ANC-DCSK systems. This can be attributed to the strong signal interferences generated by the cross product of the different user data signals and the reference signal present in the decision variable for the PNC-DCSK and ANC-DCSK schemes. Additionally, it can be seen that the PNC scheme outperforms the ANC one since the relay in the ANC scheme amplifies and forwards the resultant noisy signal to the users, which increases the multipath and noise interference in the signal. 

\begin{figure}[htbp]
\centering 
\includegraphics[width=\linewidth]{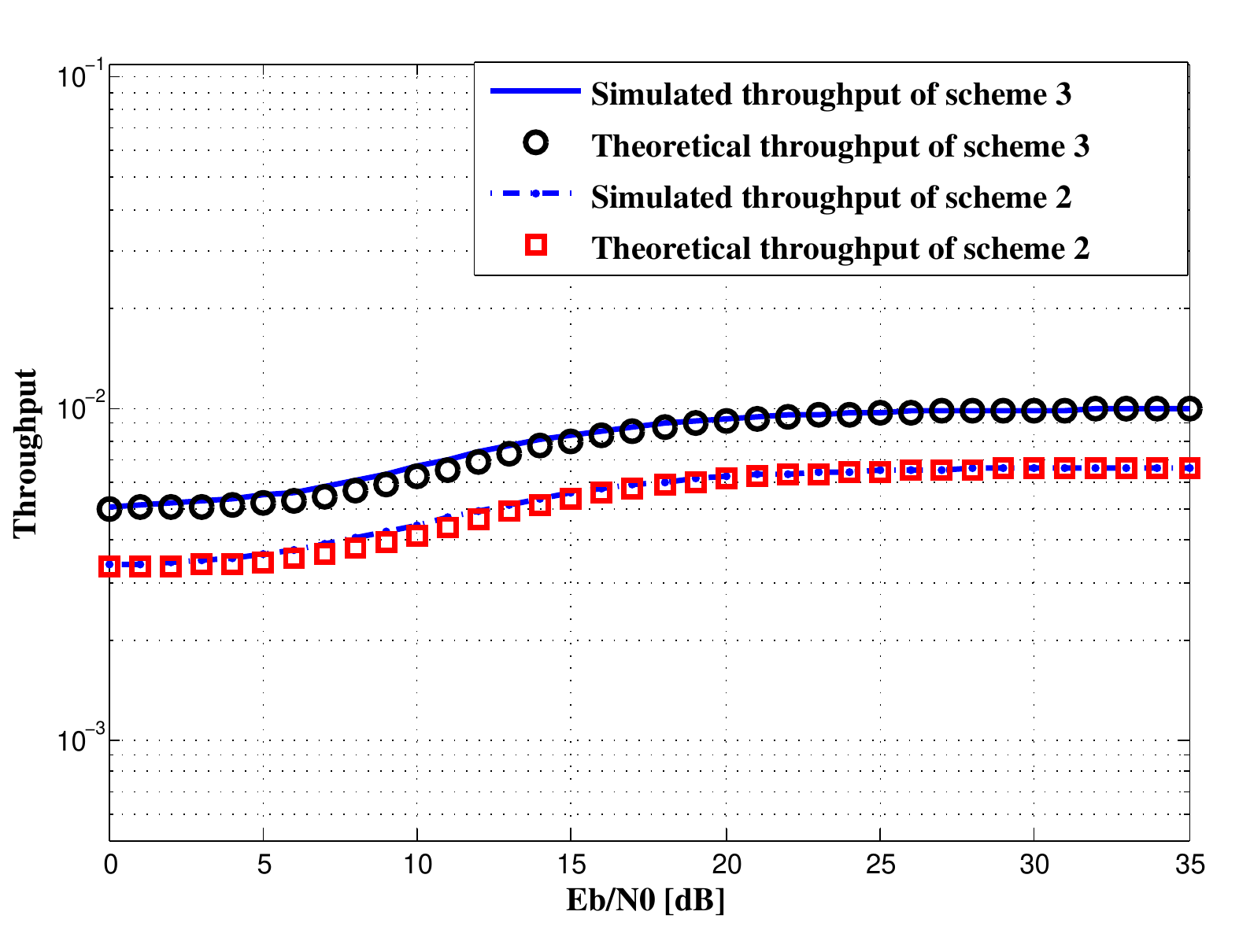}
\caption {Throughput comparison between the time multiplexed network coding scheme $2$ and the frequency multiplexed network coding scheme $3$ for $\beta=50$.
\label{throughput_comp}} 
\end{figure}
\begin{figure}[htbp]
\centering 
\includegraphics[width=\linewidth]{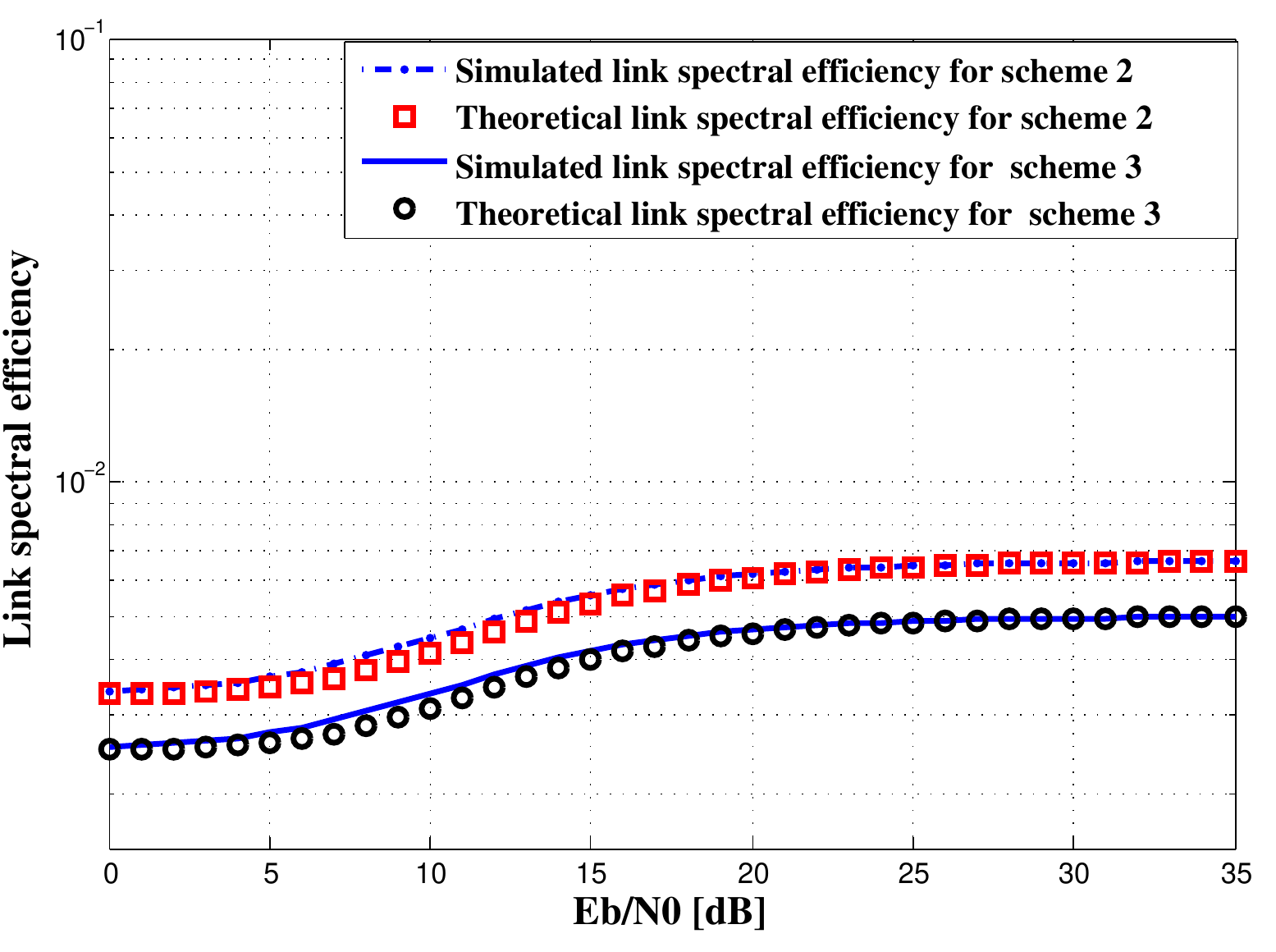}
\caption {Link spectral efficiency comparison between the time multiplexed network coding scheme $2$ and the frequency multiplexed network coding scheme $3$ for $\beta=50$.
\label{link_eff}} 
\end{figure}

Figs. \ref{throughput_comp} and \ref{link_eff} compare the throughput and the link spectral efficiency of the proposed schemes $2$ and $3$. The results in the two figures validate the obtained theoretical expressions, where the analytical results and the simulation results of the two schemes follow each other closely. As shown in these two figures, the frequency multiplexed coding scheme $3$ offers $1.5$ times higher throughput than the time multiplexed coding scheme $2$, while its link spectral efficiency is $0.75$ lower.

%
%

\section{Conclusion}
Three network coding schemes combined with DCSK modulation under a multipath fading channel were proposed and analyzed. In the proposed schemes, the users communicate through a relay node, where the relay decodes the received signals sent by different users and retransmits the mapped symbols. On the other hand, each user decodes the received symbols from the relay and recovers the data transmitted by the other user of its pair by subtracting its own data signal. The proposed scheme $1$ corresponds to the design of a PNC-DCSK scheme, where the users transmit their data synchronously to the relay while sharing the same spreading code and bandwidth. We have presented the analysis of this system showing the existence of high power interference in the combined signal at the relay. Hence, in order to mitigate this interference, we proposed to separate the transmitted signals in the first phase of relaying, where we proposed two novel schemes, which we referred to as scheme 2 and scheme 3, based on time and frequency multiplexing techniques. The separation between signals in scheme 2, which is equivalent to SNC, is performed in time domain during the first phase of relaying. On the other hand, in scheme $3$ we used frequency multiplexing to separate the two users, which requires twice the bandwidth of scheme $1$ and $2$. The performance of scheme $2$ and $3$ was analysed in different scenarios and the corresponding end-to-end bit error rate expressions under multipath Rayleigh fading channel were obtained, validated by simulations and compared to the PNC and ANC schemes. In addition, the throughput and the link spectral efficiency for the proposed schemes were derived and compared with the simulation results. 
Finally, according to the analysis of this work, the PNC-DCSK scheme $1$ will be a potential topic to study as future work. In this sense, mitigating the interference in this system leads to a significant increase in the throughput of the system, and thus considerably improving its overall performance.

\bibliographystyle{IEEEtran}
\bibliography{bibliographie_twir}

\end{document}